\documentclass[twocolumn,superscriptaddress]{revtex4-1}
\usepackage{graphicx}
\usepackage{dcolumn}
\begin{document}
\title{Valley Dynamics of Excitons in Monolayer Dichalcogenides}
\author{Gerd Plechinger}
\affiliation{Institut f\"ur Experimentelle und Angewandte Physik,
	Universit\"at Regensburg, D-93040 Regensburg, Germany}
\author{Philipp Nagler}
\affiliation{Institut f\"ur Experimentelle und Angewandte Physik,
	Universit\"at Regensburg, D-93040 Regensburg, Germany}
\author{Ashish Arora}
\affiliation{Physikalisches Institut, Westf\"alische Wilhelms-Universit\"at M\"unster, D-48149 M\"unster, Germany}
\author{Robert Schmidt}
\affiliation{Physikalisches Institut, Westf\"alische Wilhelms-Universit\"at M\"unster, D-48149 M\"unster, Germany}
\author{Alexey Chernikov}
\affiliation{Institut f\"ur Experimentelle und Angewandte Physik,
	Universit\"at Regensburg, D-93040 Regensburg, Germany}
\author{John Lupton}
\affiliation{Institut f\"ur Experimentelle und Angewandte Physik,
	Universit\"at Regensburg, D-93040 Regensburg, Germany}
\author{Rudolf Bratschitsch}
\affiliation{Physikalisches Institut, Westf\"alische Wilhelms-Universit\"at M\"unster, D-48149 M\"unster, Germany}
\author{Christian Sch\"uller}
\affiliation{Institut f\"ur Experimentelle und Angewandte Physik,
	Universit\"at Regensburg, D-93040 Regensburg, Germany}
\author{Tobias Korn}
\affiliation{Institut f\"ur Experimentelle und Angewandte Physik,
	Universit\"at Regensburg, D-93040 Regensburg, Germany}
\begin{abstract}
Monolayer transition-metal dichalcogenides (TMDCs) have recently emerged as possible candidates for valleytronic applications, as the spin and valley pseudospin are directly coupled and stabilized by a large spin splitting. In these semiconducting materials, optically excited electron-hole pairs form tightly Coulomb-bound excitons with large binding energies. The  selection rules for excitonic transitions  allow for direct optical generation of a valley-polarized exciton population using resonant excitation.
Here, we investigate the exciton valley dynamics in monolayers of three different TMDCs  by means of time-resolved Kerr rotation at low temperatures. We observe pronounced differences in the valley dynamics of tungsten- and molybdenum-based TMDCs, which are directly related to the opposite order of the conduction-band spin splitting in these materials.
\end{abstract}
\maketitle
\section{Introduction}
In recent years, the study of two-dimensional crystals has evolved into one of the most active fields in solid-state physics. Initially driven by the fascinating properties of graphene~\cite{Novoselov666}, research has rapidly expanded towards the plethora of other layered materials~\cite{C4CS00102H} that can be exfoliated into atomically thin sheets. Among these, transition-metal dichalcogenides (TMDCs) such as MoS$_2$, MoSe$_2$, WS$_2$ and WSe$_2$ are currently under intense study, as they have many interesting properties, including a large band gap that becomes direct in the monolayer limit~\cite{Mak2010,Splendiani2010}, leading to pronounced photoluminescence~\cite{Tonndorf2013}, and tightly bound excitonic states with binding energies of several hundred meV~\cite{Chernikov2014}. The band extrema are located at the K points at the corners of the hexagonal Brillouin zone, and there is a pronounced, valley-contrasting spin-splitting of conduction and valence bands at the inequivalent K$^+$ and K$^-$ points~\cite{Xiao2012}. Depending on the material composition, this spin splitting amounts to several hundreds of meV in the valence band~\cite{PhysRevB.86.115409,Kormanyos2015}, giving rise to two characteristic, spectrally well-separated excitonic transitions involving the upper (A exciton) and lower (B exciton) valence band~\cite{Mak2010,Splendiani2010}. The spin splitting in the conduction band, although being substantially smaller, was calculated to be on the order of tens of meV, and its magnitude, as well as its sign, varies substantially between molybdenum- and tungsten-based TMDCs~\cite{Kormanyos2015}. The optical selection rules in the TMDCs~\cite{Xiao2012} allow for valley-selective excitation at K$^+$ or K$^-$ using circularly polarized light, so that a valley polarization can be established by near-resonant excitation and read out via the circular polarization degree of the emitted photoluminescence~\cite{Zeng2012,Mak2012,Sallen2012}. This valley polarization might be utilized in novel optoelectronic devices which exploit the valley degree of freedom for information processing and storage (valleytronics), but to explore the potential for applications, a profound understanding of the spin and valley dynamics and the relevant dephasing mechanisms is needed. This need sparked a large number of experimental and theoretical studies on spin and valley dynamics in TMDC monolayers over the past few years. Studies based on helicity- and time-resolved photoluminescence~\cite{Lagarde2014,Wang2014c}  proved to be very challenging due to the ultrafast photocarrier dynamics in the TMDC monolayers~\cite{Korn2011,Poellmann2015,Wang2015e,Marie16}, so that alternative techniques that had been well-developed to study semiconductor spin dynamics were applied to the TMDCs, such as helicity-resolved pump-probe spectroscopy~\cite{Wang2013a,Mai2014,Mai2014a,Brat16,PhysRevLett.117.257402} and time-resolved Kerr rotation (TRKR)~\cite{Zhu2014b,Yang2015a,Plechinger16,Bushong_Arxiv,Plechinger17,Beschoten_Arxiv}. Remarkably, in these experiments, the reported values for spin and valley lifetimes range from a few picoseconds for excitons in exfoliated TMDC monolayers to nanoseconds for resident carriers in highly-doped CVD-grown materials and, most recently, hundreds of nanoseconds for optically dark, charged excitons, indicating a pronounced influence of the specific material system and excitation conditions on the valley dynamics in TMDCs.

Here, we investigate the exciton valley dynamics in mechanically exfoliated monolayers of three different TMDCs.  Using TRKR, we observe pronounced differences in the valley dynamics of tungsten- and molybdenum-based TMDCs, which are directly related to the opposite order of the spin splitting in these materials. Under A-exciton-resonant excitation, we find a very rapid decay of the Kerr signal for the tungsten-based TMDCs, followed by a long-lived tail, which we associate with the formation of dark excitons. This long-lived tail is absent in the Kerr traces measured on MoS$_2$, as the lowest-energy A exciton state is optically bright. In temperature-dependent measurements, we observe a suppression of the long-lived dark exciton population in WS$_2$ above 60~K, indicating thermal activation of the bright exciton state.  B-exciton-resonant excitation in MoS$_2$ leads to more rapid decay of valley polarization than A-exciton-resonant excitation due to the availability of additional conduction- and valence-band relaxation channels. In stark contrast, we observe a very long-lived Kerr signal under B-exciton-resonant excitation in WS$_2$, consistent with dark exciton formation or  transfer of valley polarization to resident carriers.
\section{Results and discussion}
\begin{figure}
	\includegraphics[width=  \linewidth]{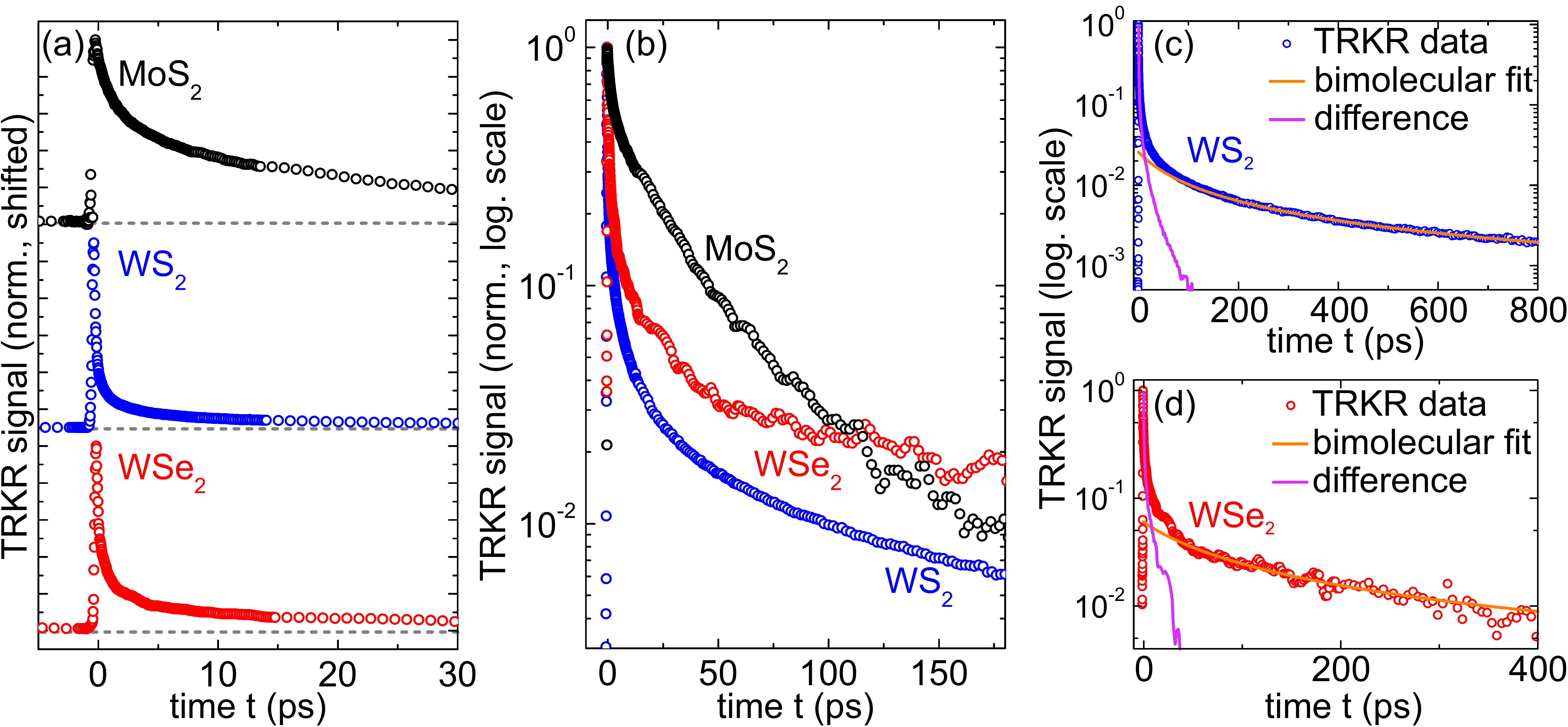}
	\caption{\label{Kerr_Exziton_Comparison}\textbf{Comparison of low-temperature A exciton valley dynamics in different TMDC monolayers.} (a) Kerr rotation traces measured on MoS$_2$, WS$_2$ and WSe$_2$ using A-exciton-resonant excitation. All traces were recorded at a nominal sample temperature of 4.5~K. (b) Same data as shown in (a) plotted on a semi-logarithmic scale. (c) and (d) TRKR traces (circles), bimolecular fits to the data (orange solid lines) and difference between data and fit (purple solid line) plotted for measurements on WS$_2$(c) and WSe$_2$(d) recorded at 4.5~K.}
\end{figure}
First, we discuss the valley dynamics in the different TMDC monolayers under A-exciton-resonant excitation at low temperature. Figure~\ref{Kerr_Exziton_Comparison} shows Kerr rotation traces measured on MoS$_2$, WS$_2$ and WSe$_2$ at a temperature of 4.5~K. For each material, the laser excitation and probe energy in the degenerate TRKR measurements was tuned to generate the maximum Kerr signal in a small energy range around the A exciton transition energy, which was independently determined using photoluminescence and white-light reflectance spectroscopy (see \cite{Plechinger16} and \cite{Plechinger17} for further details). Comparing the TRKR traces on the few-ps timescale, we find that the valley dynamics in neither of the three materials can be described by a simple exponential decay. In fact,  a triple exponential decay function is necessary to match the dynamics of the TRKR trace in MoS$_2$, and the tungsten-based materials show even more complex dynamics,  where a bimolecular decay is needed in addition to a triple exponential to fully describe the dynamics. Qualitatively, we notice that all traces show a very rapid initial decay, followed by more slowly decaying dynamics. This initial decay is on the order of the time resolution of our setup for WS$_2$ and WSe$_2$ (about 200~fs), and on the order of a picosecond for MoS$_2$. To interpret the TRKR measurements we present here, we need to consider that the Kerr rotation signals we measure are proportional to an excitonic valley polarization, i.e., the population imbalance between excitons in the K$^+$ and K$^-$ valleys. Thus, any process that reduces the exciton population, such as radiative recombination, will also reduce the Kerr signal. This is in stark contrast to measurements of the valley polarization \emph{degree} based on helicity-resolved analysis of excitonic PL emission, where the ratio between population imbalance and total population is detected. Thus, we tentatively assign this initial decay to radiative recombination of excitons within the light cone. For WSe$_2$, it was recently shown~\cite{Poellmann2015} that this radiative lifetime is about 150~fs, in good agreement with the data we extract for both tungsten-based materials. For MoS$_2$, the slightly slower initial decay indicates a smaller oscillator strength. Even though we excite our samples resonantly, exciton-phonon and exciton-exciton interactions  rapidly lead to a population of excitons at higher-\emph{k} states outside of the light cone, for which radiative recombination is suppressed. The initial decay of the Kerr traces related to radiative recombination is followed by a second decay process, which occurs on the 2-3~ps timescale for all three materials. We can assign this process to trion formation, as all of our samples are unintentionally doped. The formation of trions in our TMDC samples is directly observable as  a pronounced trion emission peak in low-temperature photoluminescence (not shown). Recently, it was  was shown for the related TMDC MoSe$_2$~\cite{Li_TrionFormation16} that trion formation occurs in about 2~ps after resonant excitation of A excitons. While trion formation should not directly influence valley depolarization, we need to consider the spectral sensitivity of our energy-degenerate Kerr setup: in the experiment, we tune the laser energy to yield a maximum Kerr rotation signal for pumping and probing the A exciton transition. As trion formation occurs, our Kerr signal is diminished, as the trion resonance is at substantially lower energy than the A exciton resonance, so that the Kerr probe becomes off-resonant.

After these processes, we note that the Kerr signal of the two tungsten-based materials decays almost completely within the following few picoseconds, while the dynamics in the MoS$_2$ are markedly slower, and a substantial Kerr signal remains in the 30~ps time window shown in Fig.~\ref{Kerr_Exziton_Comparison}(a). In order to understand these qualitative differences between the materials, we need to consider the different conduction-band spin splittings for tungsten- and molybdenum-based TMDCs~\cite{Dery_FlexPhonons,Kormanyos2015} and the corresponding energy splitting between bright and dark exciton states~\cite{Dery_Excitons,Marie_DarkBright16,Dery_intervalley}. For WS$_2$ and WSe$_2$, the optically allowed interband transition from the upper valence band, corresponding to  A-exciton-resonant excitation,  addresses the \emph{upper} conduction band. By contrast, in MoS$_2$, the \emph{lower} conduction band is addressed in the A exciton transition. Thus, in the tungsten-based TMDC monolayers, there are relaxation channels for electrons bound into A excitons, which are absent in MoS$_2$: electrons may relax into the lower spin-split subband either via spin-conserving intervalley scattering processes towards the opposite K or the $\Lambda$ valley~\cite{Selig16}, or via intravalley spin-flip transitions. All of these scattering processes result in the formation of dark excitons, which cannot recombine radiatively due to lack of momentum conservation or violation of optical selection rules. While we cannot differentiate between the different relaxation processes in our TRKR measurements, we can tentatively assign the observed decay constants of about 6~ps for WS$_2$ and 14~ps for WSe$_2$ to the formation of dark exciton populations. These dark exciton states still yield a finite Kerr signal due to the remaining valence-band occupation imbalance.
Remarkably, we find clear evidence of this subset of dark excitons in the Kerr rotation data measured for longer time delays, as shown in Fig.~\ref{Kerr_Exziton_Comparison}(b), where the TRKR traces are shown on a logarithmic scale. For both tungsten-based materials, we observe a change of the initially fast decay dynamics towards a long-lived tail of the Kerr signal at longer time delays, whereas such a change is absent in the MoS$_2$ trace. Instead, the decay of the MoS$_2$ signal on these longer time scales can be described by an exponential decay with a time constant of about 35~ps.
The long-lived tail in the Kerr traces can be well-described by a bimolecular fit function for, both, WS$_2$ and WSe$_2$, as demonstrated in Fig.~\ref{Kerr_Exziton_Comparison}(c) and (d). In both cases, the differences (purple solid lines)  between the measured data (open circles) and the bimolecular fit functions (orange solid lines) decay towards zero within the first 100~ps after excitation. The bimolecular dynamics observed in the long-lived tail are characteristic of an exciton-exciton mediated decay indicative of Auger recombination, which was shown to be efficient in TMDC monolayers~\cite{Poellmann2015}. For the MoS$_2$, there are two limiting processes that contribute to the decay of the Kerr signal on longer timescales: (a) radiative recombination of excitons that are scattered back into the light cone and (b) exciton valley depolarization driven by long-range exchange interaction~\cite{Glazov2014}.
\begin{figure}
	\includegraphics[width=  \linewidth]{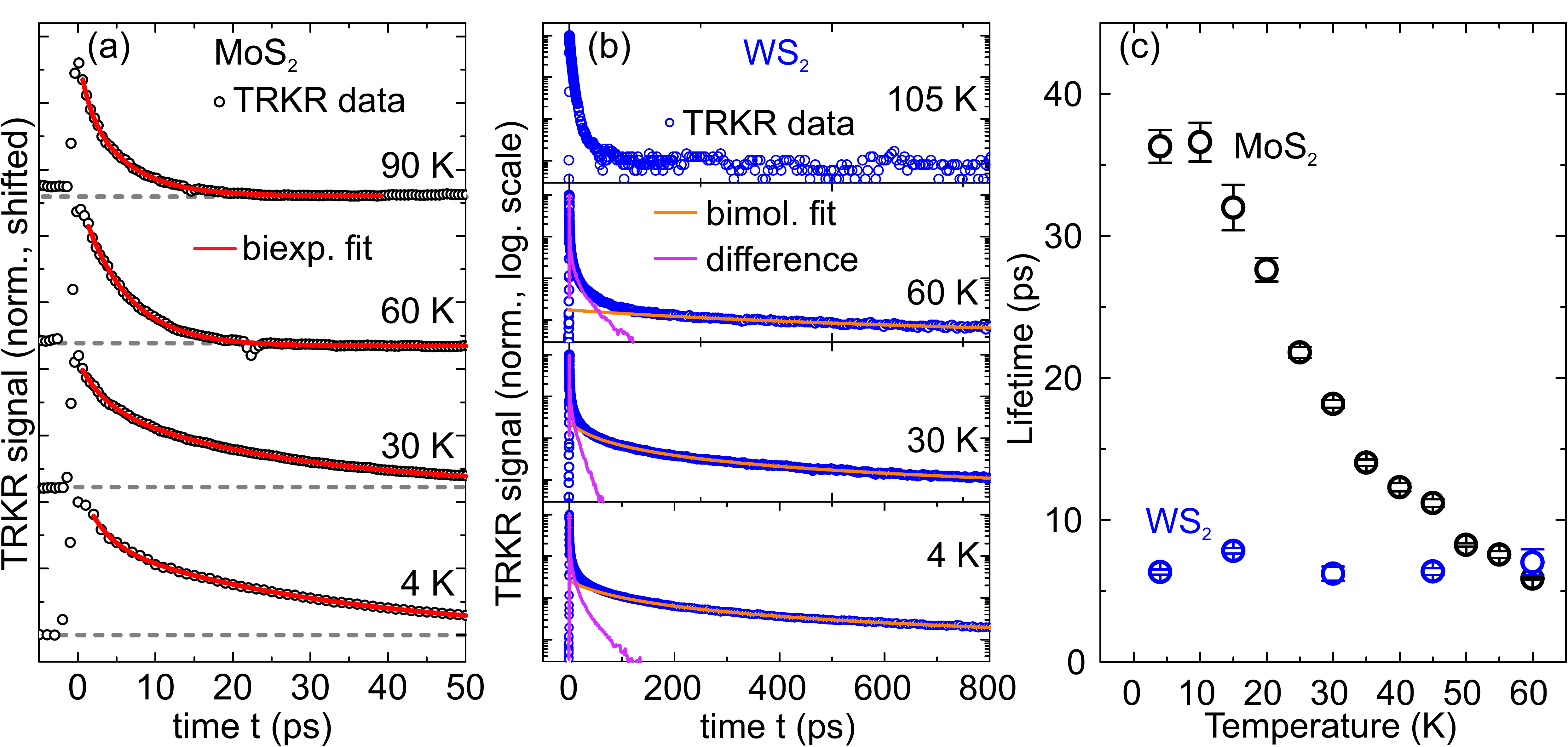}
	\caption{\label{Kerr_Temp_Dependence}\textbf{Temperature dependence of A exciton valley dynamics in MoS$_2$ and WS$_2$.}(a)  Kerr rotation traces (black circles) measured on MoS$_2$ at different sample temperatures. The red solid lines indicate biexponential fits to the data. (b) Kerr rotation traces (blue circles) measured on WS$_2$ at different sample temperatures. The  orange solid lines indicate bimolecular fits to the data, the purple solid lines show the difference between data and fit. (c) Decay constants extracted from temperature-dependent TRKR traces. For MoS$_2$, the long-lived component of the biexponential fit is plotted. For WS$_2$, the long-lived component of a triple exponential fit to the difference between TRKR trace and bimolecular fit is plotted. }
\end{figure}

Further evidence of the qualitative differences in the valley dynamics of tungsten- and molybdenum-based TMDC monolayers is provided by temperature-dependent TRKR measurements performed on MoS$_2$ and WS$_2$. In these series, the laser excitation energy was systematically red-shifted with increasing temperature to match the decreasing A exciton transition energies.  In the measurements on MoS$_2$, we used a reduced time resolution, so that the initial decay associated with radiative recombination is not well-resolved in that data.  As Fig.~\ref{Kerr_Temp_Dependence}(a) shows, the decay of the TRKR traces in MoS$_2$ is well-described by a biexponential fit, in which we tentatively assign the fast component to the trion formation dynamics, while the  long-lived component corresponds to the limiting processes discussed above. We find that with increasing temperature, the trion formation time remains almost constant at about 3~ps, while the second component of the decay becomes faster, monotonously decreasing from more than 30~ps to less than 10~ps in the range between 4.5~K and 60~K, as shown in Fig.~\ref{Kerr_Temp_Dependence}(c). This decrease may be attributed to an increasing occupation of higher-momentum exciton states, leading to an increasing valley depolarization rate~\cite{Glazov2014}.
By contrast, for WS$_2$, the decay of the TRKR traces is well-described by a combination of a  triple exponential decay and a bimolecular fit to the data in a temperature range between 4.5~K and 60~K. In this temperature range, we find that the time constant associated with dark exciton formation remains nearly constant, whereas the 'long-tail' fraction of the signal is diminished.  As the temperature is increased further, the long tail is no longer observable. This behavior is consistent with a thermal activation of dark excitons, which was previously observed to occur at a temperature of about 100~K for WSe$_2$ monolayers~\cite{Heinz_DarkEx_PRL15,Arora15,Bayer16}.
\begin{figure}
	\includegraphics[width=  \linewidth]{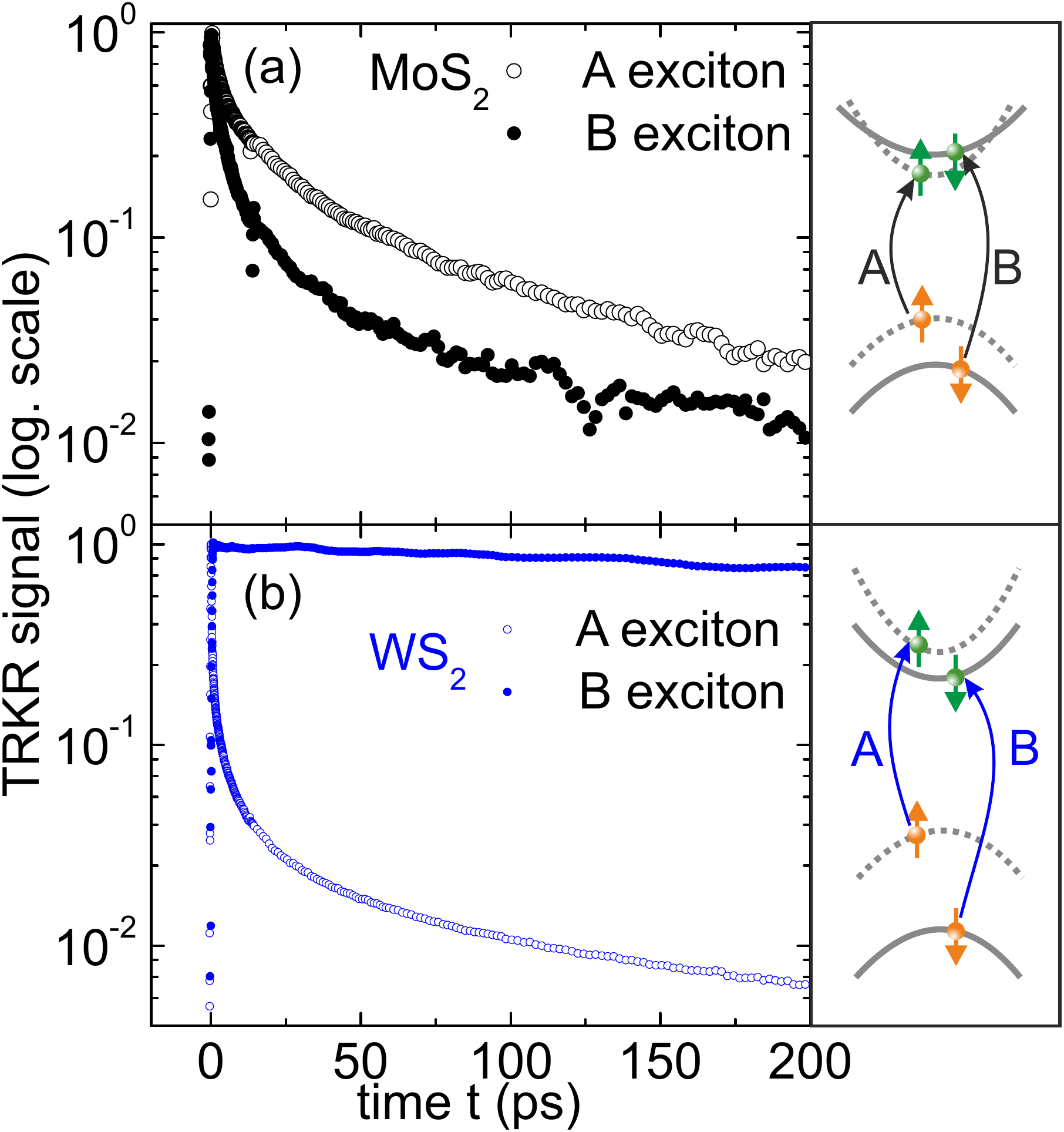}
	\caption{\label{Kerr_A_B}\textbf{A and B exciton valley dynamics in MoS$_2$ and WS$_2$.}(a)  Kerr rotation traces  measured on MoS$_2$ for A-exciton-resonant (black circles) and B-exciton-resonant (black dots) excitation. (b)  Kerr rotation traces  measured on WS$_2$ for A-exciton-resonant (blue circles)  and B-exciton-resonant (blue dots) excitation. In the panels on the right, the relevant transitions for A and B excitons in MoS$_2$ and WS$_2$ are indicated schematically.}
\end{figure}

The different spin splitting in tungsten- and molybdenum-based TMDCs should also influence the valley dynamics for B-exciton-resonant excitation. In WS$_2$ and WSe$_2$, the B exciton transition addresses the \emph{lower} conduction band, while in MoS$_2$, it addresses the \emph{upper} conduction band. The differences in these transitions are schematically shown in Fig.~\ref{Kerr_A_B}. In order to probe the consequences of this difference, we performed TRKR measurements using B-exciton-resonant excitation in WS$_2$ and MoS$_2$. For both materials, we determined the optimum excitation conditions by tuning the laser system in a small energy range around the B exciton transition energy, which was extracted from white-light reflectance measurements in the case of WS$_2$ and PL measurements in the case of MoS$_2$.  Figure~\ref{Kerr_A_B}(a) shows a direct comparison of A- and B-exciton-resonant TRKR traces measured on MoS$_2$ in a semi-logarithmic scale. We clearly see that the signal decays more rapidly under B-exciton-resonant excitation. This behavior is expected, since there are additional relaxation channels for electrons and holes towards the energetically lower conduction- and valence-band states after B-exciton-resonant excitation.

Measurements on WS$_2$ yield very different results, however, as Fig.~\ref{Kerr_A_B}(b) shows. While A-exciton-resonant excitation leads to the complex dynamics discussed above, with a rapid, multi-exponential decay followed by the long tail corresponding to the dark exciton population, the Kerr signal for B-exciton-resonant excitation shows only a very slow decay, with a $\frac{1}{e}$ time of more than 700~ps. We note that, due to the finite time window of the measurement, we cannot clearly verify whether an exponential decay or another functional dependence is more suitable to describe  the Kerr trace. Remarkably, we do not observe any rapid radiative decay of the signal within the first few ps after excitation. This is in agreement with the fact that there is no discernible PL emission from B excitons in our WS$_2$ samples, even under high-power, pulsed excitation~\cite{Plechinger16}. To explain the very unusual slow decay of the Kerr signal, we can consider different scenarios. On the one hand, a hole bound in the B exciton state can relax into the upper valence band, either in a spin-conserving intervalley scattering or in a spin-flip intravalley scattering process. In either case, this results in the formation of a dark A exciton, which would be governed by the same dynamics as discussed above for A-exciton-resonant excitation in tungsten-based TMDCs. On the other hand, valley polarization of resident carriers may occur: the B-exciton resonant excitation addresses the lower conduction band, which is partially occupied by resident electrons in our unintentionally doped samples, so that  a transfer of valley polarization via the B exciton transition could be far more efficient than via the A exciton transition. Such a transfer would require a loss of the hole valley polarization during relaxation into the upper valence band. These unpolarized holes could then recombine with electrons in either valley. Similar transfer processes, which hinge on an imbalance of spin dephasing for electrons and holes, have been observed in various GaAs-based heterostructures~\cite{PhysRevLett.102.167402,Korn2010415}. In either of these two scenarios, the Kerr signal would stem from the valley occupation imbalance in the lower conduction band, which is probed via the B exciton transition. However, in either case, we would expect an initial decay of the Kerr signal due to the relaxation and/or recombination of valley-polarized holes, in analogy to the initial decay observed for A-exciton-resonant excitation in WS$_2$. This process might be masked by the limited time resolution of our experiment, and the ultrafast dynamics under B-exciton-resonant excitation deserves further study.
\section{Conclusion}
In conclusion, we have investigated exciton valley dynamics in monolayers of three different TMDCs at low temperatures using time-resolved Kerr rotation. For tungsten-based materials, we observe  a long-tail component of the Kerr signal under A-exciton-resonant excitation, which we attribute to a population of optically dark excitons that is formed due to relaxation of electrons into the lower conduction band. This long-lived component is absent in MoS$_2$, which has an opposite conduction-band spin splitting,  supporting our interpretation.  In temperature-dependent measurements, we find that the valley polarization decay rate   in MoS$_2$ monotonously increases with temperature. In WS$_2$,  the long-lived dark exciton population is suppressed for temperatures above 60~K, indicating thermal activation of dark excitons.  Comparison of valley dynamics under A- and B-exciton-resonant excitation shows a more rapid decay for MoS$_2$ as the higher-energy B exciton is excited, due to the availability of additional conduction- and valence-band relaxation channels. By contrast, in WS$_2$, B-exciton-resonant excitation results in a very long-lived Kerr signal, which might either stem from dark exciton formation or from transfer of valley polarization to resident carriers.
\section{Methods}
MoS$_2$, WS$_2$ and WSe$_2$ flakes were prepared from  bulk crystals using an  all-dry transfer technique~\cite{Castellanos2014}, with p-doped silicon  covered with a 285\,nm SiO$_2$ layer as final substrate. For WS$_2$ and WSe$_2$, synthetic bulk crystals from HQgraphene were used, while MoS$_2$ was prepared using natural bulk crystals. Typical lateral sizes of the monolayer flakes exceed 50~$\mu$m, allowing for optical measurements with a large focal spot.
Two different pulsed laser sources were used for the time-resolved Kerr rotation (TRKR)  measurements. Measurements on MoS$_2$ and WS$_2$ were performed with a tunable, frequency-doubled pulsed fiber laser system (Toptica TVIS, pulse length (FWHM) 180~fs, spectral linewidth (FWHM) 7~meV). For WSe$_2$, a frequency-tunable Ti:sapphire laser (Chameleon Ultra II, Coherent, pulse length (FWHM) 130~fs, spectral linewidth (FWHM) 10~meV) was used.
For the TRKR measurements, the laser beam of the  laser source was split into pump and probe pulse trains, and a variable time delay between the pulses was generated using a mechanical delay stage. Pump and probe beams were focussed onto the sample to a spot size of about 30\,$\mu$m with an achromatic lens. The samples were mounted in vacuum on the cold finger of a He-flow cryostat, which could be moved beneath the fixed beam path of the TRKR experiment to position the laser spot onto a TMDC flake. For this alignment procedure, a
digital microscope was used. The pump beam was circularly polarized using an achromatic quarter-wave plate, while the probe beam was linearly polarized. A lock-in scheme, in which the intensity of the pump beam was modulated by a flywheel chopper, was employed to increase the sensitivity of the measurements. To detect changes of the reflected probe beam polarization state, an optical bridge detector was employed. For all TRKR measurements, two traces with opposite pump helicities were recorded in sequence, and the difference of these traces was calculated. This scheme ensures that any time-resolved signal that is not helicity-dependent cancels out.
\section*{Additional information}
\subsection*{Acknowledgements}
Financial support by the A. v. Humboldt foundation and the DFG via GRK 1570, KO3612/1-1,CH 1672/1-1 and SFB689 are  gratefully acknowledged.

\begin{thebibliography}{42}%
\makeatletter
\providecommand \@ifxundefined [1]{%
 \@ifx{#1\undefined}
}%
\providecommand \@ifnum [1]{%
 \ifnum #1\expandafter \@firstoftwo
 \else \expandafter \@secondoftwo
 \fi
}%
\providecommand \@ifx [1]{%
 \ifx #1\expandafter \@firstoftwo
 \else \expandafter \@secondoftwo
 \fi
}%
\providecommand \natexlab [1]{#1}%
\providecommand \enquote  [1]{``#1''}%
\providecommand \bibnamefont  [1]{#1}%
\providecommand \bibfnamefont [1]{#1}%
\providecommand \citenamefont [1]{#1}%
\providecommand \href@noop [0]{\@secondoftwo}%
\providecommand \href [0]{\begingroup \@sanitize@url \@href}%
\providecommand \@href[1]{\@@startlink{#1}\@@href}%
\providecommand \@@href[1]{\endgroup#1\@@endlink}%
\providecommand \@sanitize@url [0]{\catcode `\\12\catcode `\$12\catcode
  `\&12\catcode `\#12\catcode `\^12\catcode `\_12\catcode `\%12\relax}%
\providecommand \@@startlink[1]{}%
\providecommand \@@endlink[0]{}%
\providecommand \url  [0]{\begingroup\@sanitize@url \@url }%
\providecommand \@url [1]{\endgroup\@href {#1}{\urlprefix }}%
\providecommand \urlprefix  [0]{URL }%
\providecommand \Eprint [0]{\href }%
\providecommand \doibase [0]{http://dx.doi.org/}%
\providecommand \selectlanguage [0]{\@gobble}%
\providecommand \bibinfo  [0]{\@secondoftwo}%
\providecommand \bibfield  [0]{\@secondoftwo}%
\providecommand \translation [1]{[#1]}%
\providecommand \BibitemOpen [0]{}%
\providecommand \bibitemStop [0]{}%
\providecommand \bibitemNoStop [0]{.\EOS\space}%
\providecommand \EOS [0]{\spacefactor3000\relax}%
\providecommand \BibitemShut  [1]{\csname bibitem#1\endcsname}%
\let\auto@bib@innerbib\@empty
\bibitem [{\citenamefont {Novoselov}\ \emph {et~al.}(2004)\citenamefont
  {Novoselov}, \citenamefont {Geim}, \citenamefont {Morozov}, \citenamefont
  {Jiang}, \citenamefont {Zhang}, \citenamefont {Dubonos}, \citenamefont
  {Grigorieva},\ and\ \citenamefont {Firsov}}]{Novoselov666}%
  \BibitemOpen
  \bibfield  {author} {\bibinfo {author} {\bibfnamefont {K.~S.}\ \bibnamefont
  {Novoselov}}, \bibinfo {author} {\bibfnamefont {A.~K.}\ \bibnamefont {Geim}},
  \bibinfo {author} {\bibfnamefont {S.~V.}\ \bibnamefont {Morozov}}, \bibinfo
  {author} {\bibfnamefont {D.}~\bibnamefont {Jiang}}, \bibinfo {author}
  {\bibfnamefont {Y.}~\bibnamefont {Zhang}}, \bibinfo {author} {\bibfnamefont
  {S.~V.}\ \bibnamefont {Dubonos}}, \bibinfo {author} {\bibfnamefont {I.~V.}\
  \bibnamefont {Grigorieva}}, \ and\ \bibinfo {author} {\bibfnamefont {A.~A.}\
  \bibnamefont {Firsov}},\ }\href {\doibase 10.1126/science.1102896} {\bibfield
   {journal} {\bibinfo  {journal} {Science}\ }\textbf {\bibinfo {volume}
  {306}},\ \bibinfo {pages} {666} (\bibinfo {year} {2004})},\ \Eprint
  {http://arxiv.org/abs/http://science.sciencemag.org/content/306/5696/666.full.pdf}
  {http://science.sciencemag.org/content/306/5696/666.full.pdf} \BibitemShut
  {NoStop}%
\bibitem [{\citenamefont {Miro}\ \emph {et~al.}(2014)\citenamefont {Miro},
  \citenamefont {Audiffred},\ and\ \citenamefont {Heine}}]{C4CS00102H}%
  \BibitemOpen
  \bibfield  {author} {\bibinfo {author} {\bibfnamefont {P.}~\bibnamefont
  {Miro}}, \bibinfo {author} {\bibfnamefont {M.}~\bibnamefont {Audiffred}}, \
  and\ \bibinfo {author} {\bibfnamefont {T.}~\bibnamefont {Heine}},\ }\href
  {\doibase 10.1039/C4CS00102H} {\bibfield  {journal} {\bibinfo  {journal}
  {Chem. Soc. Rev.}\ }\textbf {\bibinfo {volume} {43}},\ \bibinfo {pages}
  {6537} (\bibinfo {year} {2014})}\BibitemShut {NoStop}%
\bibitem [{\citenamefont {Mak}\ \emph {et~al.}(2010)\citenamefont {Mak},
  \citenamefont {Lee}, \citenamefont {Hone}, \citenamefont {Shan},\ and\
  \citenamefont {Heinz}}]{Mak2010}%
  \BibitemOpen
  \bibfield  {author} {\bibinfo {author} {\bibfnamefont {K.~F.}\ \bibnamefont
  {Mak}}, \bibinfo {author} {\bibfnamefont {C.}~\bibnamefont {Lee}}, \bibinfo
  {author} {\bibfnamefont {J.}~\bibnamefont {Hone}}, \bibinfo {author}
  {\bibfnamefont {J.}~\bibnamefont {Shan}}, \ and\ \bibinfo {author}
  {\bibfnamefont {T.~F.}\ \bibnamefont {Heinz}},\ }\href {\doibase
  10.1103/PhysRevLett.105.136805} {\bibfield  {journal} {\bibinfo  {journal}
  {Physical Review Letters}\ }\textbf {\bibinfo {volume} {105}},\ \bibinfo
  {pages} {136805} (\bibinfo {year} {2010})}\BibitemShut {NoStop}%
\bibitem [{\citenamefont {Splendiani}\ \emph {et~al.}(2010)\citenamefont
  {Splendiani}, \citenamefont {Sun}, \citenamefont {Zhang}, \citenamefont {Li},
  \citenamefont {Kim}, \citenamefont {Chim}, \citenamefont {Galli},\ and\
  \citenamefont {Wang}}]{Splendiani2010}%
  \BibitemOpen
  \bibfield  {author} {\bibinfo {author} {\bibfnamefont {A.}~\bibnamefont
  {Splendiani}}, \bibinfo {author} {\bibfnamefont {L.}~\bibnamefont {Sun}},
  \bibinfo {author} {\bibfnamefont {Y.}~\bibnamefont {Zhang}}, \bibinfo
  {author} {\bibfnamefont {T.}~\bibnamefont {Li}}, \bibinfo {author}
  {\bibfnamefont {J.}~\bibnamefont {Kim}}, \bibinfo {author} {\bibfnamefont
  {C.~Y.}\ \bibnamefont {Chim}}, \bibinfo {author} {\bibfnamefont
  {G.}~\bibnamefont {Galli}}, \ and\ \bibinfo {author} {\bibfnamefont
  {F.}~\bibnamefont {Wang}},\ }\href {\doibase 10.1021/nl903868w} {\bibfield
  {journal} {\bibinfo  {journal} {Nano Letters}\ }\textbf {\bibinfo {volume}
  {10}},\ \bibinfo {pages} {1271} (\bibinfo {year} {2010})}\BibitemShut
  {NoStop}%
\bibitem [{\citenamefont {Tonndorf}\ \emph {et~al.}(2013)\citenamefont
  {Tonndorf}, \citenamefont {Schmidt}, \citenamefont {B\"{o}ttger},
  \citenamefont {Zhang}, \citenamefont {B\"{o}rner}, \citenamefont {Liebig},
  \citenamefont {Albrecht}, \citenamefont {Kloc}, \citenamefont {Gordan},
  \citenamefont {Zahn}, \citenamefont {de~Vasconcellos},\ and\ \citenamefont
  {Bratschitsch}}]{Tonndorf2013}%
  \BibitemOpen
  \bibfield  {author} {\bibinfo {author} {\bibfnamefont {P.}~\bibnamefont
  {Tonndorf}}, \bibinfo {author} {\bibfnamefont {R.}~\bibnamefont {Schmidt}},
  \bibinfo {author} {\bibfnamefont {P.}~\bibnamefont {B\"{o}ttger}}, \bibinfo
  {author} {\bibfnamefont {X.}~\bibnamefont {Zhang}}, \bibinfo {author}
  {\bibfnamefont {J.}~\bibnamefont {B\"{o}rner}}, \bibinfo {author}
  {\bibfnamefont {A.}~\bibnamefont {Liebig}}, \bibinfo {author} {\bibfnamefont
  {M.}~\bibnamefont {Albrecht}}, \bibinfo {author} {\bibfnamefont
  {C.}~\bibnamefont {Kloc}}, \bibinfo {author} {\bibfnamefont {O.}~\bibnamefont
  {Gordan}}, \bibinfo {author} {\bibfnamefont {D.~R.~T.}\ \bibnamefont {Zahn}},
  \bibinfo {author} {\bibfnamefont {S.~M.}\ \bibnamefont {de~Vasconcellos}}, \
  and\ \bibinfo {author} {\bibfnamefont {R.}~\bibnamefont {Bratschitsch}},\
  }\href@noop {} {\bibfield  {journal} {\bibinfo  {journal} {Optics Express}\
  }\textbf {\bibinfo {volume} {21}},\ \bibinfo {pages} {4908 } (\bibinfo {year}
  {2013})}\BibitemShut {NoStop}%
\bibitem [{\citenamefont {Chernikov}\ \emph {et~al.}(2014)\citenamefont
  {Chernikov}, \citenamefont {Berkelbach}, \citenamefont {Hill}, \citenamefont
  {Rigosi}, \citenamefont {Li}, \citenamefont {Aslan}, \citenamefont
  {Reichman}, \citenamefont {Hybertsen},\ and\ \citenamefont
  {Heinz}}]{Chernikov2014}%
  \BibitemOpen
  \bibfield  {author} {\bibinfo {author} {\bibfnamefont {A.}~\bibnamefont
  {Chernikov}}, \bibinfo {author} {\bibfnamefont {T.~C.}\ \bibnamefont
  {Berkelbach}}, \bibinfo {author} {\bibfnamefont {H.~M.}\ \bibnamefont
  {Hill}}, \bibinfo {author} {\bibfnamefont {A.}~\bibnamefont {Rigosi}},
  \bibinfo {author} {\bibfnamefont {Y.}~\bibnamefont {Li}}, \bibinfo {author}
  {\bibfnamefont {O.~B.}\ \bibnamefont {Aslan}}, \bibinfo {author}
  {\bibfnamefont {D.~R.}\ \bibnamefont {Reichman}}, \bibinfo {author}
  {\bibfnamefont {M.~S.}\ \bibnamefont {Hybertsen}}, \ and\ \bibinfo {author}
  {\bibfnamefont {T.~F.}\ \bibnamefont {Heinz}},\ }\href {\doibase
  10.1103/PhysRevLett.113.076802} {\bibfield  {journal} {\bibinfo  {journal}
  {Physical Review Letters}\ }\textbf {\bibinfo {volume} {113}},\ \bibinfo
  {pages} {076802} (\bibinfo {year} {2014})}\BibitemShut {NoStop}%
\bibitem [{\citenamefont {Xiao}\ \emph {et~al.}(2012)\citenamefont {Xiao},
  \citenamefont {Liu}, \citenamefont {Feng}, \citenamefont {Xu},\ and\
  \citenamefont {Yao}}]{Xiao2012}%
  \BibitemOpen
  \bibfield  {author} {\bibinfo {author} {\bibfnamefont {D.}~\bibnamefont
  {Xiao}}, \bibinfo {author} {\bibfnamefont {G.~B.}\ \bibnamefont {Liu}},
  \bibinfo {author} {\bibfnamefont {W.}~\bibnamefont {Feng}}, \bibinfo {author}
  {\bibfnamefont {X.}~\bibnamefont {Xu}}, \ and\ \bibinfo {author}
  {\bibfnamefont {W.}~\bibnamefont {Yao}},\ }\href {\doibase
  10.1103/PhysRevLett.108.196802} {\bibfield  {journal} {\bibinfo  {journal}
  {Physical Review Letters}\ }\textbf {\bibinfo {volume} {108}},\ \bibinfo
  {pages} {196802} (\bibinfo {year} {2012})}\BibitemShut {NoStop}%
\bibitem [{\citenamefont {Ramasubramaniam}(2012)}]{PhysRevB.86.115409}%
  \BibitemOpen
  \bibfield  {author} {\bibinfo {author} {\bibfnamefont {A.}~\bibnamefont
  {Ramasubramaniam}},\ }\href {\doibase 10.1103/PhysRevB.86.115409} {\bibfield
  {journal} {\bibinfo  {journal} {Phys. Rev. B}\ }\textbf {\bibinfo {volume}
  {86}},\ \bibinfo {pages} {115409} (\bibinfo {year} {2012})}\BibitemShut
  {NoStop}%
\bibitem [{\citenamefont {Korm\'{a}nyos}\ \emph {et~al.}(2015)\citenamefont
  {Korm\'{a}nyos}, \citenamefont {Burkard}, \citenamefont {Gmitra},
  \citenamefont {Fabian}, \citenamefont {Z\'{o}lyomi}, \citenamefont
  {Drummond},\ and\ \citenamefont {Fal'ko}}]{Kormanyos2015}%
  \BibitemOpen
  \bibfield  {author} {\bibinfo {author} {\bibfnamefont {A.}~\bibnamefont
  {Korm\'{a}nyos}}, \bibinfo {author} {\bibfnamefont {G.}~\bibnamefont
  {Burkard}}, \bibinfo {author} {\bibfnamefont {M.}~\bibnamefont {Gmitra}},
  \bibinfo {author} {\bibfnamefont {J.}~\bibnamefont {Fabian}}, \bibinfo
  {author} {\bibfnamefont {V.}~\bibnamefont {Z\'{o}lyomi}}, \bibinfo {author}
  {\bibfnamefont {N.~D.}\ \bibnamefont {Drummond}}, \ and\ \bibinfo {author}
  {\bibfnamefont {V.~I.}\ \bibnamefont {Fal'ko}},\ }\href {\doibase
  10.1088/2053-1583/2/2/022001} {\bibfield  {journal} {\bibinfo  {journal} {2D
  Materials}\ }\textbf {\bibinfo {volume} {2}},\ \bibinfo {pages} {022001}
  (\bibinfo {year} {2015})}\BibitemShut {NoStop}%
\bibitem [{\citenamefont {Zeng}\ \emph {et~al.}(2012)\citenamefont {Zeng},
  \citenamefont {Dai}, \citenamefont {Yao}, \citenamefont {Xiao},\ and\
  \citenamefont {Cui}}]{Zeng2012}%
  \BibitemOpen
  \bibfield  {author} {\bibinfo {author} {\bibfnamefont {H.}~\bibnamefont
  {Zeng}}, \bibinfo {author} {\bibfnamefont {J.}~\bibnamefont {Dai}}, \bibinfo
  {author} {\bibfnamefont {W.}~\bibnamefont {Yao}}, \bibinfo {author}
  {\bibfnamefont {D.}~\bibnamefont {Xiao}}, \ and\ \bibinfo {author}
  {\bibfnamefont {X.}~\bibnamefont {Cui}},\ }\href {\doibase
  10.1038/nnano.2012.95} {\bibfield  {journal} {\bibinfo  {journal} {Nature
  Nanotechnology}\ }\textbf {\bibinfo {volume} {7}},\ \bibinfo {pages} {490}
  (\bibinfo {year} {2012})}\BibitemShut {NoStop}%
\bibitem [{\citenamefont {Mak}\ \emph {et~al.}(2012)\citenamefont {Mak},
  \citenamefont {He}, \citenamefont {Shan},\ and\ \citenamefont
  {Heinz}}]{Mak2012}%
  \BibitemOpen
  \bibfield  {author} {\bibinfo {author} {\bibfnamefont {K.~F.}\ \bibnamefont
  {Mak}}, \bibinfo {author} {\bibfnamefont {K.}~\bibnamefont {He}}, \bibinfo
  {author} {\bibfnamefont {J.}~\bibnamefont {Shan}}, \ and\ \bibinfo {author}
  {\bibfnamefont {T.~F.}\ \bibnamefont {Heinz}},\ }\href@noop {} {\bibfield
  {journal} {\bibinfo  {journal} {Nature Nanotechnology}\ }\textbf {\bibinfo
  {volume} {7}},\ \bibinfo {pages} {494} (\bibinfo {year} {2012})}\BibitemShut
  {NoStop}%
\bibitem [{\citenamefont {Sallen}\ \emph {et~al.}(2012)\citenamefont {Sallen},
  \citenamefont {Bouet}, \citenamefont {Marie}, \citenamefont {Wang},
  \citenamefont {Zhu}, \citenamefont {Han}, \citenamefont {Lu}, \citenamefont
  {Tan}, \citenamefont {Amand}, \citenamefont {Liu},\ and\ \citenamefont
  {Urbaszek}}]{Sallen2012}%
  \BibitemOpen
  \bibfield  {author} {\bibinfo {author} {\bibfnamefont {G.}~\bibnamefont
  {Sallen}}, \bibinfo {author} {\bibfnamefont {L.}~\bibnamefont {Bouet}},
  \bibinfo {author} {\bibfnamefont {X.}~\bibnamefont {Marie}}, \bibinfo
  {author} {\bibfnamefont {G.}~\bibnamefont {Wang}}, \bibinfo {author}
  {\bibfnamefont {C.~R.}\ \bibnamefont {Zhu}}, \bibinfo {author} {\bibfnamefont
  {W.~P.}\ \bibnamefont {Han}}, \bibinfo {author} {\bibfnamefont
  {Y.}~\bibnamefont {Lu}}, \bibinfo {author} {\bibfnamefont {P.~H.}\
  \bibnamefont {Tan}}, \bibinfo {author} {\bibfnamefont {T.}~\bibnamefont
  {Amand}}, \bibinfo {author} {\bibfnamefont {B.~L.}\ \bibnamefont {Liu}}, \
  and\ \bibinfo {author} {\bibfnamefont {B.}~\bibnamefont {Urbaszek}},\ }\href
  {\doibase 10.1103/PhysRevB.86.081301} {\bibfield  {journal} {\bibinfo
  {journal} {Physical Review B}\ }\textbf {\bibinfo {volume} {86}},\ \bibinfo
  {pages} {081301(R)} (\bibinfo {year} {2012})}\BibitemShut {NoStop}%
\bibitem [{\citenamefont {Lagarde}\ \emph {et~al.}(2014)\citenamefont
  {Lagarde}, \citenamefont {Bouet}, \citenamefont {Marie}, \citenamefont {Zhu},
  \citenamefont {Liu}, \citenamefont {Amand}, \citenamefont {Tan},\ and\
  \citenamefont {Urbaszek}}]{Lagarde2014}%
  \BibitemOpen
  \bibfield  {author} {\bibinfo {author} {\bibfnamefont {D.}~\bibnamefont
  {Lagarde}}, \bibinfo {author} {\bibfnamefont {L.}~\bibnamefont {Bouet}},
  \bibinfo {author} {\bibfnamefont {X.}~\bibnamefont {Marie}}, \bibinfo
  {author} {\bibfnamefont {C.~R.}\ \bibnamefont {Zhu}}, \bibinfo {author}
  {\bibfnamefont {B.~L.}\ \bibnamefont {Liu}}, \bibinfo {author} {\bibfnamefont
  {T.}~\bibnamefont {Amand}}, \bibinfo {author} {\bibfnamefont {P.~H.}\
  \bibnamefont {Tan}}, \ and\ \bibinfo {author} {\bibfnamefont
  {B.}~\bibnamefont {Urbaszek}},\ }\href {\doibase
  10.1103/PhysRevLett.112.047401} {\bibfield  {journal} {\bibinfo  {journal}
  {Physical Review Letters}\ }\textbf {\bibinfo {volume} {112}},\ \bibinfo
  {pages} {047401} (\bibinfo {year} {2014})}\BibitemShut {NoStop}%
\bibitem [{\citenamefont {Wang}\ \emph {et~al.}(2014)\citenamefont {Wang},
  \citenamefont {Bouet}, \citenamefont {Lagarde}, \citenamefont {Vidal},
  \citenamefont {Balocchi}, \citenamefont {Amand}, \citenamefont {Marie},\ and\
  \citenamefont {Urbaszek}}]{Wang2014c}%
  \BibitemOpen
  \bibfield  {author} {\bibinfo {author} {\bibfnamefont {G.}~\bibnamefont
  {Wang}}, \bibinfo {author} {\bibfnamefont {L.}~\bibnamefont {Bouet}},
  \bibinfo {author} {\bibfnamefont {D.}~\bibnamefont {Lagarde}}, \bibinfo
  {author} {\bibfnamefont {M.}~\bibnamefont {Vidal}}, \bibinfo {author}
  {\bibfnamefont {a.}~\bibnamefont {Balocchi}}, \bibinfo {author}
  {\bibfnamefont {T.}~\bibnamefont {Amand}}, \bibinfo {author} {\bibfnamefont
  {X.}~\bibnamefont {Marie}}, \ and\ \bibinfo {author} {\bibfnamefont
  {B.}~\bibnamefont {Urbaszek}},\ }\href {\doibase 10.1103/PhysRevB.90.075413}
  {\bibfield  {journal} {\bibinfo  {journal} {Physical Review B}\ }\textbf
  {\bibinfo {volume} {90}},\ \bibinfo {pages} {075413} (\bibinfo {year}
  {2014})},\ \Eprint {http://arxiv.org/abs/1402.6009} {arXiv:1402.6009}
  \BibitemShut {NoStop}%
\bibitem [{\citenamefont {Korn}\ \emph {et~al.}(2011)\citenamefont {Korn},
  \citenamefont {Heydrich}, \citenamefont {Hirmer}, \citenamefont
  {Schmutzler},\ and\ \citenamefont {Sch\"{u}ller}}]{Korn2011}%
  \BibitemOpen
  \bibfield  {author} {\bibinfo {author} {\bibfnamefont {T.}~\bibnamefont
  {Korn}}, \bibinfo {author} {\bibfnamefont {S.}~\bibnamefont {Heydrich}},
  \bibinfo {author} {\bibfnamefont {M.}~\bibnamefont {Hirmer}}, \bibinfo
  {author} {\bibfnamefont {J.}~\bibnamefont {Schmutzler}}, \ and\ \bibinfo
  {author} {\bibfnamefont {C.}~\bibnamefont {Sch\"{u}ller}},\ }\href {\doibase
  10.1063/1.3636402} {\bibfield  {journal} {\bibinfo  {journal} {Applied
  Physics Letters}\ }\textbf {\bibinfo {volume} {99}},\ \bibinfo {pages} {2011}
  (\bibinfo {year} {2011})}\BibitemShut {NoStop}%
\bibitem [{\citenamefont {Poellmann}\ \emph {et~al.}(2015)\citenamefont
  {Poellmann}, \citenamefont {Steinleitner}, \citenamefont {Leierseder},
  \citenamefont {Nagler}, \citenamefont {Plechinger}, \citenamefont {Porer},
  \citenamefont {Bratschitsch}, \citenamefont {Sch\"{u}ller}, \citenamefont
  {Korn},\ and\ \citenamefont {Huber}}]{Poellmann2015}%
  \BibitemOpen
  \bibfield  {author} {\bibinfo {author} {\bibfnamefont {C.}~\bibnamefont
  {Poellmann}}, \bibinfo {author} {\bibfnamefont {P.}~\bibnamefont
  {Steinleitner}}, \bibinfo {author} {\bibfnamefont {U.}~\bibnamefont
  {Leierseder}}, \bibinfo {author} {\bibfnamefont {P.}~\bibnamefont {Nagler}},
  \bibinfo {author} {\bibfnamefont {G.}~\bibnamefont {Plechinger}}, \bibinfo
  {author} {\bibfnamefont {M.}~\bibnamefont {Porer}}, \bibinfo {author}
  {\bibfnamefont {R.}~\bibnamefont {Bratschitsch}}, \bibinfo {author}
  {\bibfnamefont {C.}~\bibnamefont {Sch\"{u}ller}}, \bibinfo {author}
  {\bibfnamefont {T.}~\bibnamefont {Korn}}, \ and\ \bibinfo {author}
  {\bibfnamefont {R.}~\bibnamefont {Huber}},\ }\href {\doibase
  10.1038/nmat4356} {\bibfield  {journal} {\bibinfo  {journal} {Nature
  Materials}\ }\textbf {\bibinfo {volume} {14}},\ \bibinfo {pages} {889}
  (\bibinfo {year} {2015})}\BibitemShut {NoStop}%
\bibitem [{\citenamefont {Wang}\ \emph {et~al.}(2015)\citenamefont {Wang},
  \citenamefont {Palleau}, \citenamefont {Amand}, \citenamefont {Tongay},
  \citenamefont {Marie},\ and\ \citenamefont {Urbaszek}}]{Wang2015e}%
  \BibitemOpen
  \bibfield  {author} {\bibinfo {author} {\bibfnamefont {G.}~\bibnamefont
  {Wang}}, \bibinfo {author} {\bibfnamefont {E.}~\bibnamefont {Palleau}},
  \bibinfo {author} {\bibfnamefont {T.}~\bibnamefont {Amand}}, \bibinfo
  {author} {\bibfnamefont {S.}~\bibnamefont {Tongay}}, \bibinfo {author}
  {\bibfnamefont {X.}~\bibnamefont {Marie}}, \ and\ \bibinfo {author}
  {\bibfnamefont {B.}~\bibnamefont {Urbaszek}},\ }\href {\doibase
  10.1063/1.4916089} {\bibfield  {journal} {\bibinfo  {journal} {Applied
  Physics Letters}\ }\textbf {\bibinfo {volume} {106}},\ \bibinfo {pages}
  {112101} (\bibinfo {year} {2015})}\BibitemShut {NoStop}%
\bibitem [{\citenamefont {Robert}\ \emph {et~al.}(2016)\citenamefont {Robert},
  \citenamefont {Lagarde}, \citenamefont {Cadiz}, \citenamefont {Wang},
  \citenamefont {Lassagne}, \citenamefont {Amand}, \citenamefont {Balocchi},
  \citenamefont {Renucci}, \citenamefont {Tongay}, \citenamefont {Urbaszek},\
  and\ \citenamefont {Marie}}]{Marie16}%
  \BibitemOpen
  \bibfield  {author} {\bibinfo {author} {\bibfnamefont {C.}~\bibnamefont
  {Robert}}, \bibinfo {author} {\bibfnamefont {D.}~\bibnamefont {Lagarde}},
  \bibinfo {author} {\bibfnamefont {F.}~\bibnamefont {Cadiz}}, \bibinfo
  {author} {\bibfnamefont {G.}~\bibnamefont {Wang}}, \bibinfo {author}
  {\bibfnamefont {B.}~\bibnamefont {Lassagne}}, \bibinfo {author}
  {\bibfnamefont {T.}~\bibnamefont {Amand}}, \bibinfo {author} {\bibfnamefont
  {A.}~\bibnamefont {Balocchi}}, \bibinfo {author} {\bibfnamefont
  {P.}~\bibnamefont {Renucci}}, \bibinfo {author} {\bibfnamefont
  {S.}~\bibnamefont {Tongay}}, \bibinfo {author} {\bibfnamefont
  {B.}~\bibnamefont {Urbaszek}}, \ and\ \bibinfo {author} {\bibfnamefont
  {X.}~\bibnamefont {Marie}},\ }\href@noop {} {\bibfield  {journal} {\bibinfo
  {journal} {Physical Review B}\ }\textbf {\bibinfo {volume} {93}},\ \bibinfo
  {pages} {205423} (\bibinfo {year} {2016})}\BibitemShut {NoStop}%
\bibitem [{\citenamefont {Wang}\ \emph {et~al.}(2013)\citenamefont {Wang},
  \citenamefont {Ge}, \citenamefont {Li}, \citenamefont {Qiu}, \citenamefont
  {Ji}, \citenamefont {Feng},\ and\ \citenamefont {Sun}}]{Wang2013a}%
  \BibitemOpen
  \bibfield  {author} {\bibinfo {author} {\bibfnamefont {Q.}~\bibnamefont
  {Wang}}, \bibinfo {author} {\bibfnamefont {S.}~\bibnamefont {Ge}}, \bibinfo
  {author} {\bibfnamefont {X.}~\bibnamefont {Li}}, \bibinfo {author}
  {\bibfnamefont {J.}~\bibnamefont {Qiu}}, \bibinfo {author} {\bibfnamefont
  {Y.}~\bibnamefont {Ji}}, \bibinfo {author} {\bibfnamefont {J.}~\bibnamefont
  {Feng}}, \ and\ \bibinfo {author} {\bibfnamefont {D.}~\bibnamefont {Sun}},\
  }\href {\doibase 10.1021/nn405419h} {\bibfield  {journal} {\bibinfo
  {journal} {ACS Nano}\ }\textbf {\bibinfo {volume} {7}},\ \bibinfo {pages}
  {11087} (\bibinfo {year} {2013})}\BibitemShut {NoStop}%
\bibitem [{\citenamefont {Mai}\ \emph {et~al.}(2014{\natexlab{a}})\citenamefont
  {Mai}, \citenamefont {Semenov}, \citenamefont {Barrette}, \citenamefont {Yu},
  \citenamefont {Jin}, \citenamefont {Cao}, \citenamefont {Kim},\ and\
  \citenamefont {Gundogdu}}]{Mai2014}%
  \BibitemOpen
  \bibfield  {author} {\bibinfo {author} {\bibfnamefont {C.}~\bibnamefont
  {Mai}}, \bibinfo {author} {\bibfnamefont {Y.~G.}\ \bibnamefont {Semenov}},
  \bibinfo {author} {\bibfnamefont {A.}~\bibnamefont {Barrette}}, \bibinfo
  {author} {\bibfnamefont {Y.}~\bibnamefont {Yu}}, \bibinfo {author}
  {\bibfnamefont {Z.}~\bibnamefont {Jin}}, \bibinfo {author} {\bibfnamefont
  {L.}~\bibnamefont {Cao}}, \bibinfo {author} {\bibfnamefont {K.~W.}\
  \bibnamefont {Kim}}, \ and\ \bibinfo {author} {\bibfnamefont
  {K.}~\bibnamefont {Gundogdu}},\ }\href {\doibase 10.1103/PhysRevB.90.041414}
  {\bibfield  {journal} {\bibinfo  {journal} {Physical Review B}\ }\textbf
  {\bibinfo {volume} {90}},\ \bibinfo {pages} {041414} (\bibinfo {year}
  {2014}{\natexlab{a}})}\BibitemShut {NoStop}%
\bibitem [{\citenamefont {Mai}\ \emph {et~al.}(2014{\natexlab{b}})\citenamefont
  {Mai}, \citenamefont {Barrette}, \citenamefont {Yu}, \citenamefont {Semenov},
  \citenamefont {Kim}, \citenamefont {Cao},\ and\ \citenamefont
  {Gundogdu}}]{Mai2014a}%
  \BibitemOpen
  \bibfield  {author} {\bibinfo {author} {\bibfnamefont {C.}~\bibnamefont
  {Mai}}, \bibinfo {author} {\bibfnamefont {A.}~\bibnamefont {Barrette}},
  \bibinfo {author} {\bibfnamefont {Y.}~\bibnamefont {Yu}}, \bibinfo {author}
  {\bibfnamefont {Y.~G.}\ \bibnamefont {Semenov}}, \bibinfo {author}
  {\bibfnamefont {K.~W.}\ \bibnamefont {Kim}}, \bibinfo {author} {\bibfnamefont
  {L.}~\bibnamefont {Cao}}, \ and\ \bibinfo {author} {\bibfnamefont
  {K.}~\bibnamefont {Gundogdu}},\ }\href {\doibase 10.1021/nl403742j}
  {\bibfield  {journal} {\bibinfo  {journal} {Nano Letters}\ }\textbf {\bibinfo
  {volume} {14}},\ \bibinfo {pages} {202} (\bibinfo {year}
  {2014}{\natexlab{b}})}\BibitemShut {NoStop}%
\bibitem [{\citenamefont {Schmidt}\ \emph {et~al.}(2016)\citenamefont
  {Schmidt}, \citenamefont {Berghäuser}, \citenamefont {Schneider},
  \citenamefont {Selig}, \citenamefont {Tonndorf}, \citenamefont {Malic},
  \citenamefont {Knorr}, \citenamefont {de~Vasconcellos},\ and\ \citenamefont
  {Bratschitsch}}]{Brat16}%
  \BibitemOpen
  \bibfield  {author} {\bibinfo {author} {\bibfnamefont {R.}~\bibnamefont
  {Schmidt}}, \bibinfo {author} {\bibfnamefont {G.}~\bibnamefont {Berghäuser}},
  \bibinfo {author} {\bibfnamefont {R.}~\bibnamefont {Schneider}}, \bibinfo
  {author} {\bibfnamefont {M.}~\bibnamefont {Selig}}, \bibinfo {author}
  {\bibfnamefont {P.}~\bibnamefont {Tonndorf}}, \bibinfo {author}
  {\bibfnamefont {E.}~\bibnamefont {Malic}}, \bibinfo {author} {\bibfnamefont
  {A.}~\bibnamefont {Knorr}}, \bibinfo {author} {\bibfnamefont {S.~M.}\
  \bibnamefont {de~Vasconcellos}}, \ and\ \bibinfo {author} {\bibfnamefont
  {R.}~\bibnamefont {Bratschitsch}},\ }\href@noop {} {\bibfield  {journal}
  {\bibinfo  {journal} {Nano Letters}\ }\textbf {\bibinfo {volume} {16}},\
  \bibinfo {pages} {2945} (\bibinfo {year} {2016})}\BibitemShut {NoStop}%
\bibitem [{\citenamefont {Singh}\ \emph
  {et~al.}(2016{\natexlab{a}})\citenamefont {Singh}, \citenamefont {Tran},
  \citenamefont {Kolarczik}, \citenamefont {Seifert}, \citenamefont {Wang},
  \citenamefont {Hao}, \citenamefont {Pleskot}, \citenamefont {Gabor},
  \citenamefont {Helmrich}, \citenamefont {Owschimikow}, \citenamefont
  {Woggon},\ and\ \citenamefont {Li}}]{PhysRevLett.117.257402}%
  \BibitemOpen
  \bibfield  {author} {\bibinfo {author} {\bibfnamefont {A.}~\bibnamefont
  {Singh}}, \bibinfo {author} {\bibfnamefont {K.}~\bibnamefont {Tran}},
  \bibinfo {author} {\bibfnamefont {M.}~\bibnamefont {Kolarczik}}, \bibinfo
  {author} {\bibfnamefont {J.}~\bibnamefont {Seifert}}, \bibinfo {author}
  {\bibfnamefont {Y.}~\bibnamefont {Wang}}, \bibinfo {author} {\bibfnamefont
  {K.}~\bibnamefont {Hao}}, \bibinfo {author} {\bibfnamefont {D.}~\bibnamefont
  {Pleskot}}, \bibinfo {author} {\bibfnamefont {N.~M.}\ \bibnamefont {Gabor}},
  \bibinfo {author} {\bibfnamefont {S.}~\bibnamefont {Helmrich}}, \bibinfo
  {author} {\bibfnamefont {N.}~\bibnamefont {Owschimikow}}, \bibinfo {author}
  {\bibfnamefont {U.}~\bibnamefont {Woggon}}, \ and\ \bibinfo {author}
  {\bibfnamefont {X.}~\bibnamefont {Li}},\ }\href {\doibase
  10.1103/PhysRevLett.117.257402} {\bibfield  {journal} {\bibinfo  {journal}
  {Phys. Rev. Lett.}\ }\textbf {\bibinfo {volume} {117}},\ \bibinfo {pages}
  {257402} (\bibinfo {year} {2016}{\natexlab{a}})}\BibitemShut {NoStop}%
\bibitem [{\citenamefont {Zhu}\ \emph {et~al.}(2014)\citenamefont {Zhu},
  \citenamefont {Zhang}, \citenamefont {Glazov}, \citenamefont {Urbaszek},
  \citenamefont {Amand}, \citenamefont {Ji}, \citenamefont {Liu},\ and\
  \citenamefont {Marie}}]{Zhu2014b}%
  \BibitemOpen
  \bibfield  {author} {\bibinfo {author} {\bibfnamefont {C.~R.}\ \bibnamefont
  {Zhu}}, \bibinfo {author} {\bibfnamefont {K.}~\bibnamefont {Zhang}}, \bibinfo
  {author} {\bibfnamefont {M.}~\bibnamefont {Glazov}}, \bibinfo {author}
  {\bibfnamefont {B.}~\bibnamefont {Urbaszek}}, \bibinfo {author}
  {\bibfnamefont {T.}~\bibnamefont {Amand}}, \bibinfo {author} {\bibfnamefont
  {Z.~W.}\ \bibnamefont {Ji}}, \bibinfo {author} {\bibfnamefont {B.~L.}\
  \bibnamefont {Liu}}, \ and\ \bibinfo {author} {\bibfnamefont
  {X.}~\bibnamefont {Marie}},\ }\href {\doibase 10.1103/PhysRevB.90.161302}
  {\bibfield  {journal} {\bibinfo  {journal} {Physical Review B}\ }\textbf
  {\bibinfo {volume} {90}},\ \bibinfo {pages} {161302} (\bibinfo {year}
  {2014})}\BibitemShut {NoStop}%
\bibitem [{\citenamefont {Yang}\ \emph {et~al.}(2015)\citenamefont {Yang},
  \citenamefont {Sinitsyn}, \citenamefont {Chen}, \citenamefont {Yuan},
  \citenamefont {Zhang}, \citenamefont {Lou},\ and\ \citenamefont
  {Crooker}}]{Yang2015a}%
  \BibitemOpen
  \bibfield  {author} {\bibinfo {author} {\bibfnamefont {L.}~\bibnamefont
  {Yang}}, \bibinfo {author} {\bibfnamefont {N.~A.}\ \bibnamefont {Sinitsyn}},
  \bibinfo {author} {\bibfnamefont {W.}~\bibnamefont {Chen}}, \bibinfo {author}
  {\bibfnamefont {J.}~\bibnamefont {Yuan}}, \bibinfo {author} {\bibfnamefont
  {J.}~\bibnamefont {Zhang}}, \bibinfo {author} {\bibfnamefont
  {J.}~\bibnamefont {Lou}}, \ and\ \bibinfo {author} {\bibfnamefont {S.~A.}\
  \bibnamefont {Crooker}},\ }\href {\doibase 10.1038/nphys3419} {\bibfield
  {journal} {\bibinfo  {journal} {Nature Physics}\ }\textbf {\bibinfo {volume}
  {11}},\ \bibinfo {pages} {830 } (\bibinfo {year} {2015})}\BibitemShut
  {NoStop}%
\bibitem [{\citenamefont {Plechinger}\ \emph {et~al.}(2016)\citenamefont
  {Plechinger}, \citenamefont {Nagler}, \citenamefont {Arora}, \citenamefont
  {Schmidt}, \citenamefont {Chernikov}, \citenamefont {del Aguila},
  \citenamefont {Christianen}, \citenamefont {R.Bratschitsch}, \citenamefont
  {Sch\"{u}ller},\ and\ \citenamefont {Korn}}]{Plechinger16}%
  \BibitemOpen
  \bibfield  {author} {\bibinfo {author} {\bibfnamefont {G.}~\bibnamefont
  {Plechinger}}, \bibinfo {author} {\bibfnamefont {P.}~\bibnamefont {Nagler}},
  \bibinfo {author} {\bibfnamefont {A.}~\bibnamefont {Arora}}, \bibinfo
  {author} {\bibfnamefont {R.}~\bibnamefont {Schmidt}}, \bibinfo {author}
  {\bibfnamefont {A.}~\bibnamefont {Chernikov}}, \bibinfo {author}
  {\bibfnamefont {A.~G.}\ \bibnamefont {del Aguila}}, \bibinfo {author}
  {\bibfnamefont {P.~C.}\ \bibnamefont {Christianen}}, \bibinfo {author}
  {\bibnamefont {R.Bratschitsch}}, \bibinfo {author} {\bibfnamefont
  {C.}~\bibnamefont {Sch\"{u}ller}}, \ and\ \bibinfo {author} {\bibfnamefont
  {T.}~\bibnamefont {Korn}},\ }\href@noop {} {\bibfield  {journal} {\bibinfo
  {journal} {Nature Communications}\ }\textbf {\bibinfo {volume} {7}},\
  \bibinfo {pages} {12715} (\bibinfo {year} {2016})}\BibitemShut {NoStop}%
\bibitem [{\citenamefont {Bushong}\ \emph {et~al.}(2016)\citenamefont
  {Bushong}, \citenamefont {Luo}, \citenamefont {McCreary}, \citenamefont
  {Newburger}, \citenamefont {Singh}, \citenamefont {Jonker},\ and\
  \citenamefont {Kawakami}}]{Bushong_Arxiv}%
  \BibitemOpen
  \bibfield  {author} {\bibinfo {author} {\bibfnamefont {E.~J.}\ \bibnamefont
  {Bushong}}, \bibinfo {author} {\bibfnamefont {Y.}~\bibnamefont {Luo}},
  \bibinfo {author} {\bibfnamefont {K.~M.}\ \bibnamefont {McCreary}}, \bibinfo
  {author} {\bibfnamefont {M.~J.}\ \bibnamefont {Newburger}}, \bibinfo {author}
  {\bibfnamefont {S.}~\bibnamefont {Singh}}, \bibinfo {author} {\bibfnamefont
  {B.~T.}\ \bibnamefont {Jonker}}, \ and\ \bibinfo {author} {\bibfnamefont
  {R.~K.}\ \bibnamefont {Kawakami}},\ }\href {http://arXiv:1602.03568}
  {\bibfield  {journal} {\bibinfo  {journal} {arXiv:1602.03568}\ } (\bibinfo
  {year} {2016})}\BibitemShut {NoStop}%
\bibitem [{\citenamefont {Plechinger}\ \emph {et~al.}(2017)\citenamefont
  {Plechinger}, \citenamefont {Korn},\ and\ \citenamefont
  {Lupton}}]{Plechinger17}%
  \BibitemOpen
  \bibfield  {author} {\bibinfo {author} {\bibfnamefont {G.}~\bibnamefont
  {Plechinger}}, \bibinfo {author} {\bibfnamefont {T.}~\bibnamefont {Korn}}, \
  and\ \bibinfo {author} {\bibfnamefont {J.~M.}\ \bibnamefont {Lupton}},\
  }\href {\doibase 10.1021/acs.jpcc.7b01468} {\bibfield  {journal} {\bibinfo
  {journal} {The Journal of Physical Chemistry C}\ }\textbf {\bibinfo {volume}
  {121}},\ \bibinfo {pages} {6409} (\bibinfo {year} {2017})}\BibitemShut
  {NoStop}%
\bibitem [{\citenamefont {Volmer}\ \emph {et~al.}(2017)\citenamefont {Volmer},
  \citenamefont {Pissinger}, \citenamefont {Ersfeld}, \citenamefont {Kuhlen},
  \citenamefont {Stampfer},\ and\ \citenamefont {Beschoten}}]{Beschoten_Arxiv}%
  \BibitemOpen
  \bibfield  {author} {\bibinfo {author} {\bibfnamefont {F.}~\bibnamefont
  {Volmer}}, \bibinfo {author} {\bibfnamefont {S.}~\bibnamefont {Pissinger}},
  \bibinfo {author} {\bibfnamefont {M.}~\bibnamefont {Ersfeld}}, \bibinfo
  {author} {\bibfnamefont {S.}~\bibnamefont {Kuhlen}}, \bibinfo {author}
  {\bibfnamefont {C.}~\bibnamefont {Stampfer}}, \ and\ \bibinfo {author}
  {\bibfnamefont {B.}~\bibnamefont {Beschoten}},\ }\href
  {https://arxiv.org/abs/1702.03712} {\bibfield  {journal} {\bibinfo  {journal}
  {arXiv:1702.03712}\ } (\bibinfo {year} {2017})}\BibitemShut {NoStop}%
\bibitem [{\citenamefont {Singh}\ \emph
  {et~al.}(2016{\natexlab{b}})\citenamefont {Singh}, \citenamefont {Moody},
  \citenamefont {Tran}, \citenamefont {Scott}, \citenamefont {Overbeck},
  \citenamefont {Bergh\"auser}, \citenamefont {Schaibley}, \citenamefont
  {Seifert}, \citenamefont {Pleskot}, \citenamefont {Gabor}, \citenamefont
  {Yan}, \citenamefont {Mandrus}, \citenamefont {Richter}, \citenamefont
  {Malic}, \citenamefont {Xu},\ and\ \citenamefont {Li}}]{Li_TrionFormation16}%
  \BibitemOpen
  \bibfield  {author} {\bibinfo {author} {\bibfnamefont {A.}~\bibnamefont
  {Singh}}, \bibinfo {author} {\bibfnamefont {G.}~\bibnamefont {Moody}},
  \bibinfo {author} {\bibfnamefont {K.}~\bibnamefont {Tran}}, \bibinfo {author}
  {\bibfnamefont {M.~E.}\ \bibnamefont {Scott}}, \bibinfo {author}
  {\bibfnamefont {V.}~\bibnamefont {Overbeck}}, \bibinfo {author}
  {\bibfnamefont {G.}~\bibnamefont {Bergh\"auser}}, \bibinfo {author}
  {\bibfnamefont {J.}~\bibnamefont {Schaibley}}, \bibinfo {author}
  {\bibfnamefont {E.~J.}\ \bibnamefont {Seifert}}, \bibinfo {author}
  {\bibfnamefont {D.}~\bibnamefont {Pleskot}}, \bibinfo {author} {\bibfnamefont
  {N.~M.}\ \bibnamefont {Gabor}}, \bibinfo {author} {\bibfnamefont
  {J.}~\bibnamefont {Yan}}, \bibinfo {author} {\bibfnamefont {D.~G.}\
  \bibnamefont {Mandrus}}, \bibinfo {author} {\bibfnamefont {M.}~\bibnamefont
  {Richter}}, \bibinfo {author} {\bibfnamefont {E.}~\bibnamefont {Malic}},
  \bibinfo {author} {\bibfnamefont {X.}~\bibnamefont {Xu}}, \ and\ \bibinfo
  {author} {\bibfnamefont {X.}~\bibnamefont {Li}},\ }\href {\doibase
  10.1103/PhysRevB.93.041401} {\bibfield  {journal} {\bibinfo  {journal}
  {Physical Review B}\ }\textbf {\bibinfo {volume} {93}},\ \bibinfo {pages}
  {041401(R)} (\bibinfo {year} {2016}{\natexlab{b}})}\BibitemShut {NoStop}%
\bibitem [{\citenamefont {Song}\ and\ \citenamefont
  {Dery}(2013)}]{Dery_FlexPhonons}%
  \BibitemOpen
  \bibfield  {author} {\bibinfo {author} {\bibfnamefont {Y.}~\bibnamefont
  {Song}}\ and\ \bibinfo {author} {\bibfnamefont {H.}~\bibnamefont {Dery}},\
  }\href {\doibase 10.1103/PhysRevLett.111.026601} {\bibfield  {journal}
  {\bibinfo  {journal} {Physical Review Letters}\ }\textbf {\bibinfo {volume}
  {111}},\ \bibinfo {pages} {026601} (\bibinfo {year} {2013})}\BibitemShut
  {NoStop}%
\bibitem [{\citenamefont {Dery}\ and\ \citenamefont
  {Song}(2015)}]{Dery_Excitons}%
  \BibitemOpen
  \bibfield  {author} {\bibinfo {author} {\bibfnamefont {H.}~\bibnamefont
  {Dery}}\ and\ \bibinfo {author} {\bibfnamefont {Y.}~\bibnamefont {Song}},\
  }\href {\doibase 10.1103/PhysRevB.92.125431} {\bibfield  {journal} {\bibinfo
  {journal} {Physical Review B}\ }\textbf {\bibinfo {volume} {92}},\ \bibinfo
  {pages} {125431} (\bibinfo {year} {2015})}\BibitemShut {NoStop}%
\bibitem [{\citenamefont {Echeverry}\ \emph {et~al.}(2016)\citenamefont
  {Echeverry}, \citenamefont {Urbaszek}, \citenamefont {Amand}, \citenamefont
  {Marie},\ and\ \citenamefont {Gerber}}]{Marie_DarkBright16}%
  \BibitemOpen
  \bibfield  {author} {\bibinfo {author} {\bibfnamefont {J.~P.}\ \bibnamefont
  {Echeverry}}, \bibinfo {author} {\bibfnamefont {B.}~\bibnamefont {Urbaszek}},
  \bibinfo {author} {\bibfnamefont {T.}~\bibnamefont {Amand}}, \bibinfo
  {author} {\bibfnamefont {X.}~\bibnamefont {Marie}}, \ and\ \bibinfo {author}
  {\bibfnamefont {I.~C.}\ \bibnamefont {Gerber}},\ }\href {\doibase
  10.1103/PhysRevB.93.121107} {\bibfield  {journal} {\bibinfo  {journal} {Phys.
  Rev. B}\ }\textbf {\bibinfo {volume} {93}},\ \bibinfo {pages} {121107}
  (\bibinfo {year} {2016})}\BibitemShut {NoStop}%
\bibitem [{\citenamefont {Dery}(2016)}]{Dery_intervalley}%
  \BibitemOpen
  \bibfield  {author} {\bibinfo {author} {\bibfnamefont {H.}~\bibnamefont
  {Dery}},\ }\href {\doibase 10.1103/PhysRevB.94.075421} {\bibfield  {journal}
  {\bibinfo  {journal} {Phys. Rev. B}\ }\textbf {\bibinfo {volume} {94}},\
  \bibinfo {pages} {075421} (\bibinfo {year} {2016})}\BibitemShut {NoStop}%
\bibitem [{\citenamefont {Selig}\ \emph {et~al.}(2016)\citenamefont {Selig},
  \citenamefont {Bergh\"{a}user}, \citenamefont {Raja}, \citenamefont {Nagler},
  \citenamefont {Sch\"{u}ller}, \citenamefont {Heinz}, \citenamefont {Korn},
  \citenamefont {A.Chernikov}, \citenamefont {Malic},\ and\ \citenamefont
  {Knorr}}]{Selig16}%
  \BibitemOpen
  \bibfield  {author} {\bibinfo {author} {\bibfnamefont {M.}~\bibnamefont
  {Selig}}, \bibinfo {author} {\bibfnamefont {G.}~\bibnamefont
  {Bergh\"{a}user}}, \bibinfo {author} {\bibfnamefont {A.}~\bibnamefont
  {Raja}}, \bibinfo {author} {\bibfnamefont {P.}~\bibnamefont {Nagler}},
  \bibinfo {author} {\bibfnamefont {C.}~\bibnamefont {Sch\"{u}ller}}, \bibinfo
  {author} {\bibfnamefont {T.~F.}\ \bibnamefont {Heinz}}, \bibinfo {author}
  {\bibfnamefont {T.}~\bibnamefont {Korn}}, \bibinfo {author} {\bibnamefont
  {A.Chernikov}}, \bibinfo {author} {\bibfnamefont {E.}~\bibnamefont {Malic}},
  \ and\ \bibinfo {author} {\bibfnamefont {A.}~\bibnamefont {Knorr}},\
  }\href@noop {} {\bibfield  {journal} {\bibinfo  {journal} {Nature
  Communications}\ }\textbf {\bibinfo {volume} {7}},\ \bibinfo {pages} {13279}
  (\bibinfo {year} {2016})}\BibitemShut {NoStop}%
\bibitem [{\citenamefont {Glazov}\ \emph {et~al.}(2014)\citenamefont {Glazov},
  \citenamefont {Amand}, \citenamefont {Marie}, \citenamefont {Lagarde},
  \citenamefont {Bouet},\ and\ \citenamefont {Urbaszek}}]{Glazov2014}%
  \BibitemOpen
  \bibfield  {author} {\bibinfo {author} {\bibfnamefont {M.~M.}\ \bibnamefont
  {Glazov}}, \bibinfo {author} {\bibfnamefont {T.}~\bibnamefont {Amand}},
  \bibinfo {author} {\bibfnamefont {X.}~\bibnamefont {Marie}}, \bibinfo
  {author} {\bibfnamefont {D.}~\bibnamefont {Lagarde}}, \bibinfo {author}
  {\bibfnamefont {L.}~\bibnamefont {Bouet}}, \ and\ \bibinfo {author}
  {\bibfnamefont {B.}~\bibnamefont {Urbaszek}},\ }\href {\doibase
  10.1103/PhysRevB.89.201302} {\bibfield  {journal} {\bibinfo  {journal}
  {Physical Review B}\ }\textbf {\bibinfo {volume} {89}},\ \bibinfo {pages}
  {201302} (\bibinfo {year} {2014})}\BibitemShut {NoStop}%
\bibitem [{\citenamefont {Zhang}\ \emph {et~al.}(2015)\citenamefont {Zhang},
  \citenamefont {You}, \citenamefont {Zhao},\ and\ \citenamefont
  {Heinz}}]{Heinz_DarkEx_PRL15}%
  \BibitemOpen
  \bibfield  {author} {\bibinfo {author} {\bibfnamefont {X.-X.}\ \bibnamefont
  {Zhang}}, \bibinfo {author} {\bibfnamefont {Y.}~\bibnamefont {You}}, \bibinfo
  {author} {\bibfnamefont {S.~Y.~F.}\ \bibnamefont {Zhao}}, \ and\ \bibinfo
  {author} {\bibfnamefont {T.~F.}\ \bibnamefont {Heinz}},\ }\href {\doibase
  10.1103/PhysRevLett.115.257403} {\bibfield  {journal} {\bibinfo  {journal}
  {Physical Review Letters}\ }\textbf {\bibinfo {volume} {115}},\ \bibinfo
  {pages} {257403} (\bibinfo {year} {2015})}\BibitemShut {NoStop}%
\bibitem [{\citenamefont {Arora}\ \emph {et~al.}(2015)\citenamefont {Arora},
  \citenamefont {Koperski}, \citenamefont {Nogajewski}, \citenamefont {Marcus},
  \citenamefont {Faugeras},\ and\ \citenamefont {Potemski}}]{Arora15}%
  \BibitemOpen
  \bibfield  {author} {\bibinfo {author} {\bibfnamefont {A.}~\bibnamefont
  {Arora}}, \bibinfo {author} {\bibfnamefont {M.}~\bibnamefont {Koperski}},
  \bibinfo {author} {\bibfnamefont {K.}~\bibnamefont {Nogajewski}}, \bibinfo
  {author} {\bibfnamefont {J.}~\bibnamefont {Marcus}}, \bibinfo {author}
  {\bibfnamefont {C.}~\bibnamefont {Faugeras}}, \ and\ \bibinfo {author}
  {\bibfnamefont {M.}~\bibnamefont {Potemski}},\ }\href {\doibase
  10.1039/C5NR01536G} {\bibfield  {journal} {\bibinfo  {journal} {Nanoscale}\
  }\textbf {\bibinfo {volume} {7}},\ \bibinfo {pages} {10421} (\bibinfo {year}
  {2015})}\BibitemShut {NoStop}%
\bibitem [{\citenamefont {Godde}\ \emph {et~al.}(2016)\citenamefont {Godde},
  \citenamefont {Schmidt}, \citenamefont {Schmutzler}, \citenamefont
  {A\ss{}mann}, \citenamefont {Debus}, \citenamefont {Withers}, \citenamefont
  {Alexeev}, \citenamefont {Del Pozo-Zamudio}, \citenamefont {Skrypka},
  \citenamefont {Novoselov}, \citenamefont {Bayer},\ and\ \citenamefont
  {Tartakovskii}}]{Bayer16}%
  \BibitemOpen
  \bibfield  {author} {\bibinfo {author} {\bibfnamefont {T.}~\bibnamefont
  {Godde}}, \bibinfo {author} {\bibfnamefont {D.}~\bibnamefont {Schmidt}},
  \bibinfo {author} {\bibfnamefont {J.}~\bibnamefont {Schmutzler}}, \bibinfo
  {author} {\bibfnamefont {M.}~\bibnamefont {A\ss{}mann}}, \bibinfo {author}
  {\bibfnamefont {J.}~\bibnamefont {Debus}}, \bibinfo {author} {\bibfnamefont
  {F.}~\bibnamefont {Withers}}, \bibinfo {author} {\bibfnamefont {E.~M.}\
  \bibnamefont {Alexeev}}, \bibinfo {author} {\bibfnamefont {O.}~\bibnamefont
  {Del Pozo-Zamudio}}, \bibinfo {author} {\bibfnamefont {O.~V.}\ \bibnamefont
  {Skrypka}}, \bibinfo {author} {\bibfnamefont {K.~S.}\ \bibnamefont
  {Novoselov}}, \bibinfo {author} {\bibfnamefont {M.}~\bibnamefont {Bayer}}, \
  and\ \bibinfo {author} {\bibfnamefont {A.~I.}\ \bibnamefont {Tartakovskii}},\
  }\href {\doibase 10.1103/PhysRevB.94.165301} {\bibfield  {journal} {\bibinfo
  {journal} {Phys. Rev. B}\ }\textbf {\bibinfo {volume} {94}},\ \bibinfo
  {pages} {165301} (\bibinfo {year} {2016})}\BibitemShut {NoStop}%
\bibitem [{\citenamefont {Yugova}\ \emph {et~al.}(2009)\citenamefont {Yugova},
  \citenamefont {Sokolova}, \citenamefont {Yakovlev}, \citenamefont {Greilich},
  \citenamefont {Reuter}, \citenamefont {Wieck},\ and\ \citenamefont
  {Bayer}}]{PhysRevLett.102.167402}%
  \BibitemOpen
  \bibfield  {author} {\bibinfo {author} {\bibfnamefont {I.~A.}\ \bibnamefont
  {Yugova}}, \bibinfo {author} {\bibfnamefont {A.~A.}\ \bibnamefont
  {Sokolova}}, \bibinfo {author} {\bibfnamefont {D.~R.}\ \bibnamefont
  {Yakovlev}}, \bibinfo {author} {\bibfnamefont {A.}~\bibnamefont {Greilich}},
  \bibinfo {author} {\bibfnamefont {D.}~\bibnamefont {Reuter}}, \bibinfo
  {author} {\bibfnamefont {A.~D.}\ \bibnamefont {Wieck}}, \ and\ \bibinfo
  {author} {\bibfnamefont {M.}~\bibnamefont {Bayer}},\ }\href {\doibase
  10.1103/PhysRevLett.102.167402} {\bibfield  {journal} {\bibinfo  {journal}
  {Phys. Rev. Lett.}\ }\textbf {\bibinfo {volume} {102}},\ \bibinfo {pages}
  {167402} (\bibinfo {year} {2009})}\BibitemShut {NoStop}%
\bibitem [{\citenamefont {Korn}(2010)}]{Korn2010415}%
  \BibitemOpen
  \bibfield  {author} {\bibinfo {author} {\bibfnamefont {T.}~\bibnamefont
  {Korn}},\ }\href {\doibase http://dx.doi.org/10.1016/j.physrep.2010.05.001}
  {\bibfield  {journal} {\bibinfo  {journal} {Physics Reports}\ }\textbf
  {\bibinfo {volume} {494}},\ \bibinfo {pages} {415 } (\bibinfo {year}
  {2010})}\BibitemShut {NoStop}%
\bibitem [{\citenamefont {Castellanos-Gomez}\ \emph {et~al.}(2014)\citenamefont
  {Castellanos-Gomez}, \citenamefont {Buscema}, \citenamefont {Molenaar},
  \citenamefont {Singh}, \citenamefont {Janssen}, \citenamefont {van~der
  Zant},\ and\ \citenamefont {Steele}}]{Castellanos2014}%
  \BibitemOpen
  \bibfield  {author} {\bibinfo {author} {\bibfnamefont {A.}~\bibnamefont
  {Castellanos-Gomez}}, \bibinfo {author} {\bibfnamefont {M.}~\bibnamefont
  {Buscema}}, \bibinfo {author} {\bibfnamefont {R.}~\bibnamefont {Molenaar}},
  \bibinfo {author} {\bibfnamefont {V.}~\bibnamefont {Singh}}, \bibinfo
  {author} {\bibfnamefont {L.}~\bibnamefont {Janssen}}, \bibinfo {author}
  {\bibfnamefont {H.~S.~J.}\ \bibnamefont {van~der Zant}}, \ and\ \bibinfo
  {author} {\bibfnamefont {G.~A.}\ \bibnamefont {Steele}},\ }\href {\doibase
  10.1088/2053-1583/1/1/011002} {\bibfield  {journal} {\bibinfo  {journal} {2D
  Materials}\ }\textbf {\bibinfo {volume} {1}},\ \bibinfo {pages} {011002}
  (\bibinfo {year} {2014})}\BibitemShut {NoStop}%
\end{thebibliography}
\end{document}